\documentclass[prb,aps,preprint,superscriptaddress,floatfix]{revtex4}
\usepackage[latin1]{inputenc}
\usepackage{graphicx}
\usepackage{float}
\usepackage{amsmath}
\usepackage{natbib}
\usepackage{esint}
\usepackage{eucal}
\usepackage{color}

\begin{document}

\title{Topological Sound}
\author{Xiujuan Zhang}
\affiliation{National Laboratory of Solid State Microstructures and Department of Materials Science and Engineering, Nanjing University, Nanjing 210093, China}
\author{Meng Xiao}
\affiliation{Department of Electrical Engineering, and Ginzton Laboratory, Stanford University, Stanford, California 94305, USA}
\author{Ying Cheng}
\affiliation{Key Laboratory of Modern Acoustics, Department of Physics and Collaborative Innovation Center of Advanced Microstructures, Nanjing University, Nanjing 210093, China}
\author{Ming-Hui Lu}
\affiliation{National Laboratory of Solid State Microstructures and Department of Materials Science and Engineering, Nanjing University, Nanjing 210093, China}
\author{Johan Christensen}
\affiliation{Department of Physics, Universidad Carlos III de Madrid, ES-28916 Legan\`es, Madrid, Spain}

\begin{abstract}
Recently, we witnessed a tremendous effort to conquer the realm of acoustics as a possible playground to test with sound waves topologically protected wave propagation. Acoustics differ substantially from photonic and electronic systems since longitudinal sound waves lack intrinsic spin polarization and breaking the time-reversal symmetry requires additional complexities that both are essential in mimicking the quantum effects leading to topologically robust sound propagation. In this article, we review the latest efforts to explore with sound waves topological states of quantum matter in two- and three-dimensional systems where we discuss how spin and valley degrees of freedom appear as highly novel ingredients to tailor the flow of sound in the form of one-way edge modes and defect-immune protected acoustic waves. Both from a theoretical stand point and based on contemporary experimental verifications, we summarize the latest advancements of the flourishing research frontier on topological sound.
\end{abstract}

\maketitle

\section{Introduction}

In condensed matter physics, the distinctive phases of matter are characterized by their underlying symmetries that are spontaneously broken. For example, in a crystal, the ions are periodically arranged, breaking the continuous symmetry of space. This is a clear signature of crystals. Using such way to classify the phases of matter remains a recurring theme until the discovery of the quantum Hall effect (QHE) \cite{klitzing1980new}. In 1980s, Von Klitzing found that a two-dimensional (2D) electron gas sample, subjected to low temperature and strong magnetic field, has a quantized Hall conductance, which is independent of sample size and immune to impurities. It was later demonstrated that the state responsible for such phenomena does not break any symmetries, but is characterized by a completely different classification paradigm based on the notion of topological order \cite{laughlin1981quantized,thouless1982quantized}, therefore opening a new research branch.

\begin{figure}[htbp]
\begin{center}
\includegraphics[width=1\columnwidth]{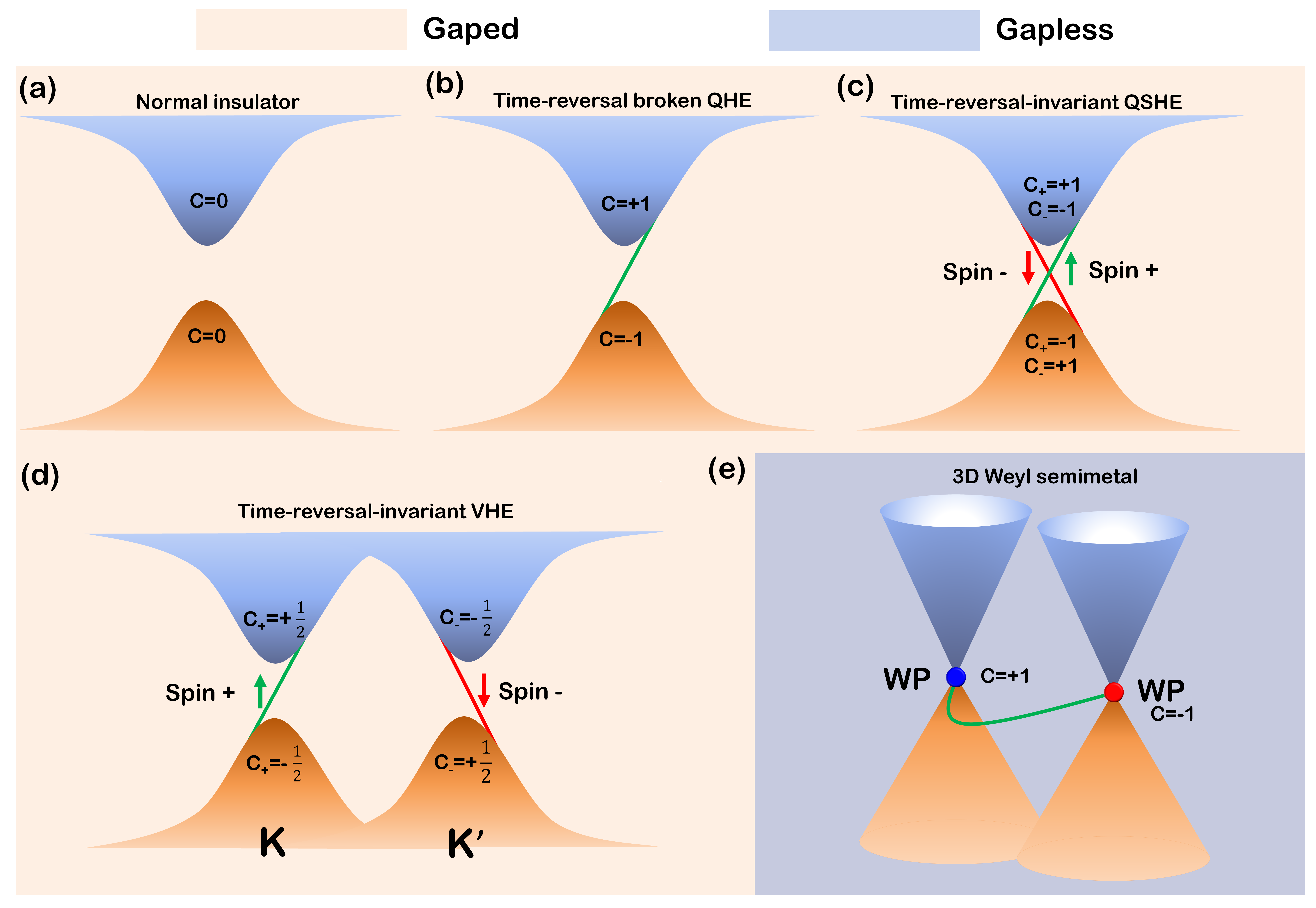}
\caption{Schematic illustration of (a) a normal insulator, (b) the time-reversal broken QHE, (c) the time-reversal-invariant QSHE, (d) the time-reversal-invariant VHE and (e) a three-dimensional (3D) Weyl semimetal. The associated topological invariants (for the gapped topological states) and the chiral charge (for the gapless 3D Weyl semimetal) are labeled for clarification.}
\label{fig:Intofig}
\end{center}
\end{figure}

The topological description on phases of matter concerns the fundamental properties of the system that are insensitive to continuous perturbations of material parameters and change only under a quantum phase transition. For the QHE, the Hall conductance is such a fundamental property. Its quantization originates from the non-trivial topological properties of the energy bands, which are featured with a non-zero topological invariant, the Chern number, according to the TKNN theory \cite{thouless1982quantized}. The Chern number characterizes the geometric phase (commonly known as the Berry phase \cite{berry1984quantal}) accumulation over the Brillouin zone, and thus is closely related with the behaviors of the energy bands in the momentum space. It has been shown that a periodic magnetic flux, which breaks the time-reversal symmetry, is able to produce non-zero Chern number \cite{haldane1988model}. The resulting topologically non-trivial system supports a gapless edge state in the bulk energy gap (Fig. \ref{fig:Intofig}(b)), exhibiting an interesting electronic property that is insulating in the bulk but conducting on the edge.  This is essentially different from a normal insulator (Fig. \ref{fig:Intofig}(a)), where the Chern number is zero.

Additional to applying a magnetic field, it was quickly found that the inherent spin-orbital coupling of a material can also give rise to non-trivial topological phases \cite{hasan2010colloquium,qi2011topological}. Kane and Zhang \textit{et al.} found in systems with spin-orbital coupling, a pair of gapless edge states emerges in the insulating bandgap. The edge states carry conjugate electronic spins and exhibit spin-dependent propagation behaviors, as sketched in Fig. \ref{fig:Intofig}(c). This is the so-called quantum spin Hall effect (QSHE). In this case, the total Hall conductance is zero, so is the Chern number, implying the time-reversal symmetry is intact. In fact, it is exactly the time-reversal symmetry that protects the spin-dependent edge states. Though the total Hall conductance is zero, the spin Hall conductance is non-zero and can be described by a $Z_2$ topological invariant or the spin Chern number \cite{sheng2005nondissipative}.

Recently, another discrete degree of freedom, namely the valley, has also been proposed to realize a topological state, known as the valley Hall effect (VHE), which is related to valleytronics \cite{yao2008valley,ju2015topological}. Valley refers to the two energy extrema of the band structure in momentum space, at which the Berry curvature exhibits opposite signs and therefore its integral over the full Brillouin zone is zero, while the integral within each valley is non-zero. As a result, the system shows a valley-selective topologically non-trivial property (Fig. \ref{fig:Intofig}(d)). It is worth mentioning that the VHE also maintains intact time-reversal symmetry. Regardless of the nature of abovementioned topological phases, they share the same property that the edge states span the bulk bandgap and separate domains with different Chern numbers. In parallel with these gapped topological phases, topological semimetals, which are featured with topologically protected gapless band structures and accompanied by gapless surface states, have recently emerged as a new frontier \cite{Burkov1,chenfang2,Burkov2}. Among them, Weyl semimetals have received particular attention, as their quasiparticle excitation is the Weyl fermion, which has not yet been observed as a fundamental particle in vacuum. In a Weyl semimetal, the Weyl points separated in momentum space carry opposite chiral charges and are connected across the domain boundaries by a surface state, i.e., the Fermi arc (Fig. \ref{fig:Intofig}(e)), upon which the Weyl fermions are robust while carrying currents.

The above mentioned topological states and their associated exotic phenomena could promise potential applications in the next generation of electronic devices and topological quantum computing. However, realizing the topological phases poses great challenges that are difficult to overcome in electronic systems, such as the inevitable material defects as well as the validity of the single electron approximation, which is the essential basis of most topological descriptions. Thus, it is not surprised that many of the quantum topological states have been extended to photonic and phononic systems, benefiting from their large scale in both time and space, which makes the control of fabrication and the measurement process much easier and more accurate compared to the electronic systems. Additionally, the photonic/phononic systems are not restricted by the Fermi levels, and therefore any appropriate regions of the spectrum can be of interest. Nevertheless, emulating the topological phases in condensed matter physics to the classical regime might not be straightforward, due to the key difference between electrons and photons/phonons. For instance, the photons/phonons do not carry a half-integer spin and therefore cannot directly interact with the magnetic field. Breaking the time-reversal symmetry in these systems requires additional effort. Moreover, the difference between fermions and bosons might also provide new angles to the quest of topological phases of matter, which may have potential applications in the design of low loss photonic/phononic devices. This review offers a detailed exposition on some of the recent advancements of the topological states in classical regime, mainly focused on the airborne sound.

The organization of the remainder of this review is as follows. In Section II, we elaborate the main breakthroughs of the analogue QHE and QSHE in acoustics, followed by the development of VHE in Section III. Section IV is devoted to the Weyl semimetal. In Section V, the possibility to extend the topological phases to mechanical waves is discussed. The last section presents our perspectives on possible future directions.

\section{Analogue quantum Hall effect and quantum spin Hall effect}

The QHE provided the first example of the topological phases of matter. Observing the QHE, a 2D sample subjected to low temperature and a strong magnetic field exhibited the behavior of an insulator in the bulk and a metal along the edges where the electrons move unidirectionally without backscattering or dissipation. The associated Hall conductance takes the quantized values of $\sigma_{xy}=Ce^2/h$, which are unaffected by impurities. Here, $h$ represents the the Plank constant, $e$ is the charge of an electron and $C$ is the Chern number. As mentioned above, it characterizes the topology of the electronic wave functions in the momentum space and is independent on the material properties \cite{laughlin1981quantized,thouless1982quantized}. This is essentially the reason why the QHE is topologically robust against impurities. For a 2D system, the Chern number can be evaluated by
\begin{equation}
C_n= \frac{1}{2\pi}\oiint F(\boldsymbol{k}) \cdot d\boldsymbol{s},
\label{eqn:Chernnumber}
\end{equation}
where $F(\boldsymbol{k})=\nabla_{\boldsymbol{k}} \times A(\boldsymbol{k})$ is defined as the Berry curvature and $A(\boldsymbol{k})=\langle u_n(\boldsymbol{k})|i\nabla_{\boldsymbol{k}}|u_n(\boldsymbol{k}) \rangle$ is the Berry connection. $u_n(\boldsymbol{k})$ represents the periodic part of the Bloch state on the $n$th energy band with momentum $\boldsymbol{k}$. Under the symmetry operations, the Berry curvature obeys the rules of $\CMcal{P}F(\boldsymbol{k})=F(-\boldsymbol{k})$ and $\CMcal{T}F(\boldsymbol{k})=-F(-\boldsymbol{k})$, where $\CMcal{P}$ and $\CMcal{T}$ denote the parity and time-reversal operators, respectively. Note that when the system breaks $\CMcal{T}$ symmetry but preserving the $\CMcal{P}$ symmetry, the integral in Eq.(\ref{eqn:Chernnumber}) (which runs over the entire Brillouin zone) acquires a non-zero value, so does the Chern number. Non-zero Chern number corresponds to a topological non-trivial phase while zero Chern number corresponds to a topological trivial phase. The topological phases with non-zero Chern number offer intriguing wave transport properties like one-way edge propagation and robustness against impurities, which might have promising applications in the next-generation of electronic devices and quantum computing.

\begin{figure}[htbp]
\begin{center}
\includegraphics[width=1\columnwidth]{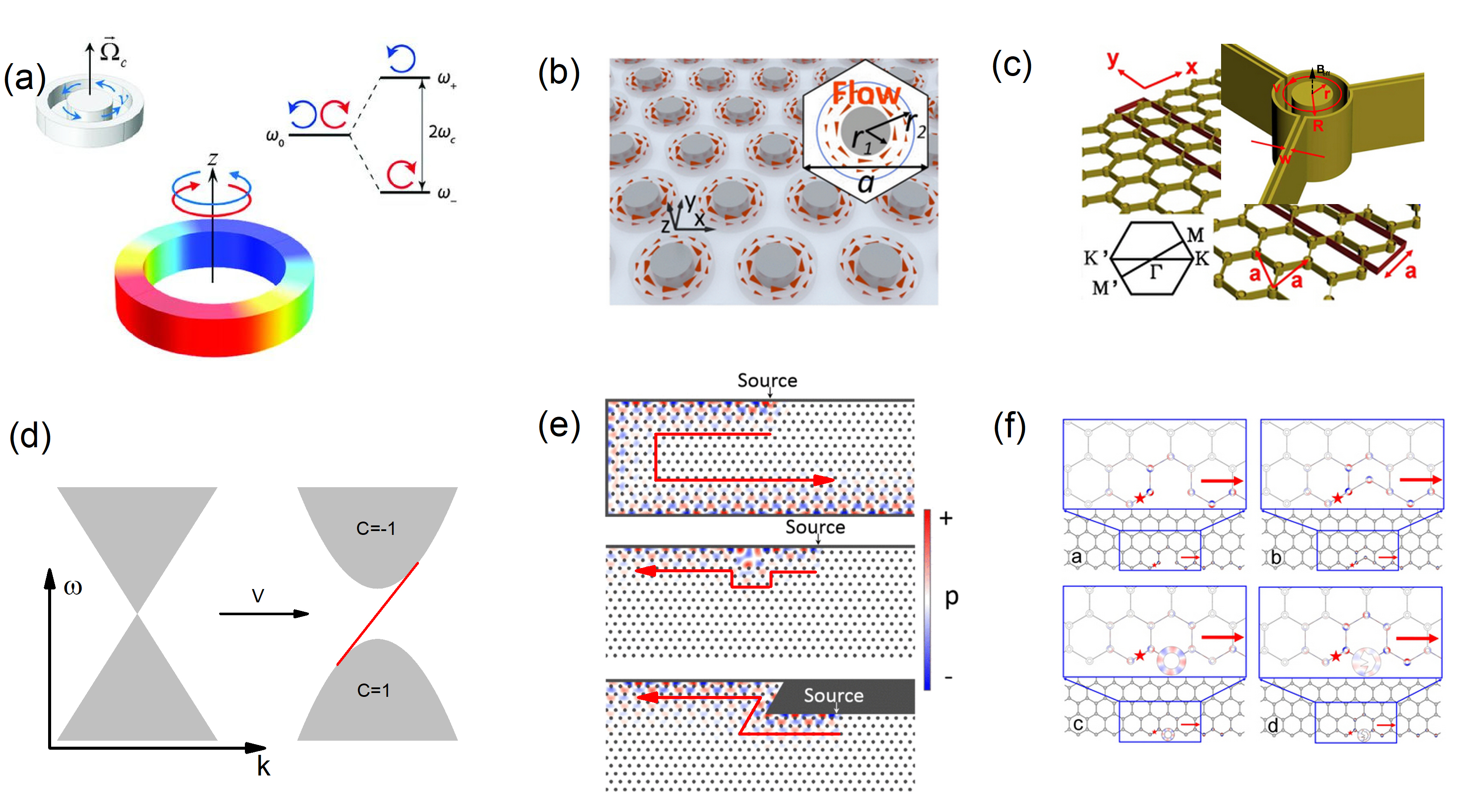}
\caption{(a) The airflow-induced acoustic nonreciprocity. In (b) and (c), two acoustic quantum Hall lattices incorporated with circulating airflow are presented. (d) Illustration of the band gap opening induced by the airflow, associated with a one-way edge state that spans the bulk gap region. (e) and (f) The robust edge state propagation against various defects. Sub-figures are adapted from: (a) Ref \cite{fleury2014sound}, AAAS; (b) and (e) Ref \cite{yang2015topological}, APS; (c) and (f) Ref \cite{ni2015topologically}, IOP.}
\label{fig:QHEfig}
\end{center}
\end{figure}

Driven by their development in electronic systems, the topological phases were quickly transferred to the classical realms, with the analogue QHE in photonics firstly proposed and experimentally realized at microwave frequencies \cite{haldane2008possible,wang2009observation}. The considered system is a 2D photonic crystal, comprising a gyromagnetic material subjected to a magnetic field that breaks the $\CMcal{T}$ symmetry. Consequently, topologically protected edge states were constructed, featured with one-way wave transport behaviors that are robust against defects and bends. In acoustics, however, breaking the $\CMcal{T}$ symmetry is quite challenging, usually involving additional complexities, such as using magneto-acoustic materials \cite{kittel1958interaction} or introducing nonlinearity \cite{liang2010acoustic}. These possibilities either require large volumes or introduce inherent signal distortion, which are typically impractical. Inspired by the magnetic bias producing electromagnetic nonreciprocity in gyromagnetic materials, Fleury \textit{et al.} proposed a feasible method to break the $\CMcal{T}$ symmetry in acoustics, relying on moving airflow in ring cavities \cite{fleury2014sound}. The imparted airflow, taking the role of magnetic bias, splits the degeneracy of the two counter-propagating azimuthal resonant modes in the ring cavities, as shown in Fig. \ref{fig:QHEfig}(a), therefore inducing acoustic nonreciprocity. More specifically, consider the following wave equation for sound propagating in a circulating air flow with velocity $\vec{V}$ \cite{ni2015topologically},
\begin{equation}
[(\nabla-i\vec{A}_{eff})^2+\omega^2/c^2+(\nabla\rho/2\rho)^2-\nabla^2\rho/(2\rho)]\phi= 0,
\label{eqn:eqinairflow}
\end{equation}
where $\phi$ is the velocity potential, $\omega$ is the angular frequency, $c$ is the sound speed and $\rho$ is the mass density of air. For non-zero $\vec{V}$, the term $\vec{A}_{eff}=-\omega\vec{V}/c^2$ gives rise to an effective vector potential, which generates an effective magnetic field $\vec{B}_{eff}=\nabla \times \vec{A}_{eff}$ that breaks the $\CMcal{T}$ symmetry. The inset in Fig. \ref{fig:QHEfig}(c) illustrates such a process.

Based on this principle, several designs of analogue QHE in sonic crystals have been reported \cite{ni2015topologically,yang2015topological,khanikaev2015topologically,chen2016tunable,chen2017acoustic}. In Figs. \ref{fig:QHEfig}(b) and \ref{fig:QHEfig}(c), a hexagonal lattice and a honeycomb lattice, respectively, are illustrated as examples. Due to the intrinsic lattice symmetry, a pair of Dirac-like points deterministically appears at the Brillouin zone boundary for $\vec{V}=0$; when the airflow is introduced, the Dirac-like degeneracies are lifted as a consequence of the $\CMcal{T}$ symmetry breaking \cite{ni2015topologically,yang2015topological}. This is characterized by a band gap opening, as illustrated in Fig. \ref{fig:QHEfig}(d). The evaluated Chern number for the bands below and above the opened gap acquires non-zero values ($C=1$ for the lower band and $C=-1$ for the upper band), implying the systems are in the topologically non-trivial phase. According to the principle of the bulk-edge correspondence, a signature of such topologically non-trivial system is the presence of one-way edge states along its boundaries when truncated by other topologically trivial systems (i.e., $C=0$). Correspondingly, acoustic wave propagation exhibits unidirectional behaviors, which are topologically protected and robust against various defects and sharp bends, as demonstrated by Figs. \ref{fig:QHEfig}(e) and \ref{fig:QHEfig}(f). 

To experimentally implement the above discussed airflow-based designs, uniformly biased circulators are required, which impose serious
challenges, such as nonsynchronous rotation and flow instability. This makes the practical implementation elusive until recently, Zhu \textit{et al.} proposed a rotating chiral structure based on ring resonators that support high-order whispering gallery modes with high Q factor \cite{zhu2018experimental}. This special design allows the system to produce giant acoustic nonreciprocity at small rotation speed, and therefore a stable and uniform airflow can be generated. On the other hand, using active liquids that can flow spontaneously even without an external drive has also been explored to break the $\CMcal{T}$ symmetry \cite{souslov2017topological,shankar2017topological}. This might relax the experimental requirements and bring new opportunities to topological phases of matter in active materials, for which, uniquely inherent material properties like microscopic irreversibility may help to achieve functionalities that are absent using only passive materials.

\begin{figure}[htbp]
\begin{center}
\includegraphics[width=0.8\columnwidth]{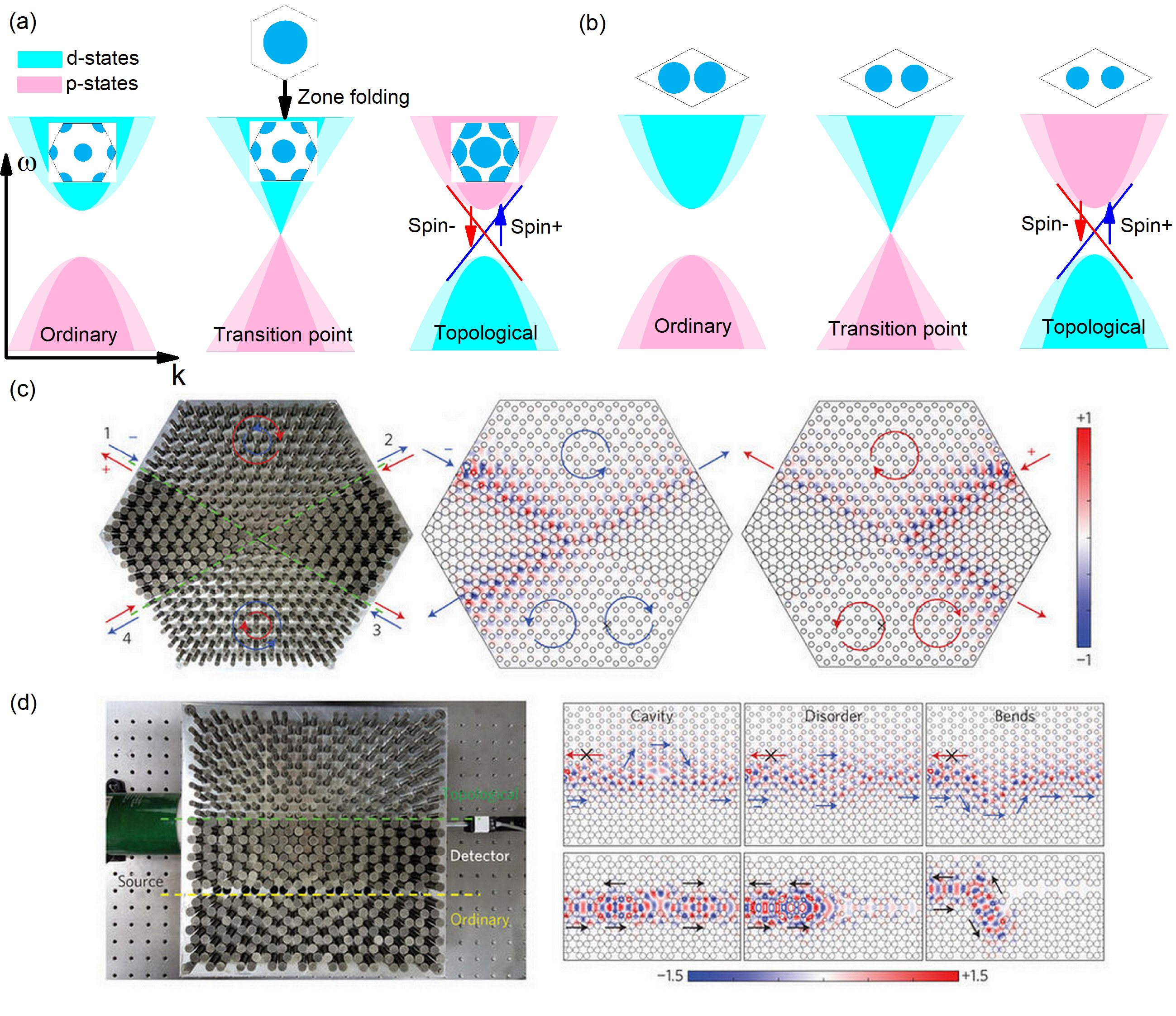}
\caption{The construction of doubly degenerate Dirac-like cones based on (a) zone folding and (b) accidental degeneracy is presented. (c) The spin-locked edge state propagation and (d) its robustness against various defects. Wave propagation in a regular waveguide is also presented as comparison (the lower panel of (d)). Panels (c) and (d) are adapted from Ref \cite{he2016acoustic}, Macmillan Publishers Ltd.}
\label{fig:QSHEfig}
\end{center}
\end{figure}

Compared to the QHE that requires to break the $\CMcal{T}$ symmetry and therefore imposes certain challenges to the experimental implementation, it is more preferable and practical to explore topological phases under preserved $\CMcal{T}$ symmetry, i.e., the QSHE, also known as the topological insulators (TIs) \cite{hasan2010colloquium,qi2011topological}. The QSHE can be regarded as the effect of two coupled quantum Hall states. Differently, the spin-orbit coupling plays an essential role in the QSHE where the coupling between spin and orbital angular momentum causes the moving electrons to feel a spin-dependent force, even in the absence of magnetic materials. As a result, the electrons with opposite spin angular momenta (often referred as spin up and spin down) will move in opposite directions along the edges. The QSHE with edge states that are spin-locked and protected by the $\CMcal{T}$ symmetry also found their counterparts in photonics and phononics, but not straightforward. It is well known that for fermions with half-integer spin, like the electrons, the $\CMcal{T}$ symmetry operator satisfies $\CMcal{T}_f^2=-1$, and hence guarantees the Kramers degeneracy, which is crucial for the QSHE \cite{murakami2004spin}. However, for bosons with integer spin, like the photons and phonons, the $\CMcal{T}$ symmetry operator obeys $\CMcal{T}_b^2=1$, which is essentially different from the fermions. Consequently, to realize analogue QSHE in bosonic systems, it is necessary to construct fermion-like pseudo spins and pseudo $\CMcal{T}$ symmetry \cite{he2016photonic}. In photonics, different polarizations were used to construct pseudo spins as TE$+$TM/TE$-$TM (where TE and TM are the transverse electric and magnetic polarizations) \cite{khanikaev2013photonic}, as TE/TM \cite{ma2015guiding} and as left/right circular polarizations \cite{he2016photonic}. In acoustics, however, due to the lack of various polarizations (sound propagates longitudinally only), it is even more challenging to realize analogue QSHE. A possible solution was addressed based on using coupled resonators that support clockwise and anticlockwise resonant modes, which impart the pseudo spins \cite{hafezi2011robust,hafezi2013imaging,zhu2018topological,peng2016experimental,he2016topological}. Recently, another approach was proposed utilizing two degenerate Bloch modes, instead of two polarizations or two resonant modes, to create the pseudo spin states \cite{wu2015scheme}. Specifically, by expanding a primitive unit cell to a larger cell, the Dirac-like cones at the
$K$ and $K'$ points in the original Brillouin zone are folded at the $\Gamma$ point in the new Brillouin zone, forming the doubly degenerate Dirac-like cones. By tuning the composite material or geometric parameters, a band inversion can happen near the degenerate point, associated with a topological transition (Fig. \ref{fig:QSHEfig}(a)). The harmonic polarized electromagnetic (EM) modes \cite{wu2015scheme} or the scalar acoustic modes \cite{zhang2017topological,yves2017topological,deng2017observation} exhibit electronic orbital $p$-like and $d$-like wave shapes, which essentially give the correspondence between the EM or acoustic wave functions with positive and negative angular momenta and the spin-up and down states of the electrons. Later on, an accidental degeneracy technique \cite{chen2014accidental} was also implemented to create the doubly degenerate Dirac-like cones and the pseudo spins \cite{he2016acoustic,mei2016pseudo}, taking the advantage of large index and impedance contrast of the composite materials, which is especially common in acoustics.

Here, we demonstrate in detail how the acoustic analogue QSHE can be realized in a honeycomb lattice consisting of steel rods in air, based on the accidental degeneracy \cite{he2016acoustic}. Due to the 2D irreducible representations of the $C_{6v}$ symmetry, the honeycomb lattice supports two pairs of degeneracies at the $\Gamma$ point, the dipolar modes $p_x$/$p_y$ and the quadrupolar modes $d_{x^2-y^2}$/$d_{xy}$, which can hybridize to emulate the pseudo spins. By decreasing the filling ratio of the steel rods, the two pairs of dipolar and quadrupolar modes, separated by a band gap, will move in frequency and exchange their positions (the so-called band inversion). In between, there is a point where the band gap is closed and the two pairs accidentally touch together, forming the doubly degenerate Dirac-like cones (essentially different from the zone folding mechanism). This gap-opened, closed and re-opened process is sketched in Fig. \ref{fig:QSHEfig}(b), which leads to a topological transition from the trivial (ordinary) state to the non-trivial (topological) state. The transition point is exactly the double Dirac-like point. In the topological non-trivial state, a pair of edge states appear, carrying opposite group velocities to emulate the spin-up and down states. Correspondingly, the spin-dependent sound propagation can be expected, which is depicted in Fig. \ref{fig:QSHEfig}(c). More studies in Fig. \ref{fig:QSHEfig}(d) reveal that the spin-locked edge state propagation is immune to various defects, including cavities, disorders and bends, essentially different from a regular waveguide (the lower penal of Fig. \ref{fig:QSHEfig}(d)). It is worth mentioning that as the $C_{6v}$ symmetry is not perfectly preserved at the interface between the trivial and non-trivial lattices, the two counter-propagating pseudo spin states are slightly mixed and a tiny gap exists at the center of the Brillouin zone. As a result, the backscattering of the edge states is not completely suppressed. Fortunately, by engineering the rod size, this gap can be sufficiently reduced. Recently, the accidental degeneracy technique also spurred the development of topological phases in the elastic realm, where an elastic analogue QSHE was experimentally demonstrated in a phononic crystal plate with perforated holes \cite{yu1707monolithic}. This might have potential applications in chip-scale acoustic devices, such as waveguide and splitter.

\section{Acoustic valley-Hall and pseudo spin effect}

The discrete valley degree of freedom \cite{NatPhys.3.172,PhysRevLett.99.236809,PhysRevLett.101.087204,PhysRevLett.100.036804,PhysRevLett.106.156801,NanoLett.11.3453,PNAS.110.10546,PhysRevX.3.021018,Science.346.448,NatPhys.10.343,Science.344.1489,ju2015topological,NatNano.11.1060}, labeling quantum states of energy extrema in momentum space, is attracting rapid growing attention because of its potential as a
new type of information carrier like spins in spintronics. Transferring the valley concept to
classical wave systems, it has been shown that the existence of valley-like frequency dispersions,
engineered in articial crystals, has been made possible with photonic crystals \cite{PhysRevLett.100.113903,PhysRevB.78.045122,PhysRevLett.104.043903,PhysRevB.82.014301,PhysRevLett.110.106801,NatMat.16.298,PhysRevB.96.020202,SciRep.8.1588,hafezi2011robust,NewJPhys.18.025012,NatPhy.14.140,SciRep.6.28453,NatComm.8.1304,PhysRevLett.120.063902,PhysRevB.96.201402,NatPhys.14.111,ApplPhysLett.111.251107} and sonic crystals (SCs) \cite{PhysRevLett.101.264303,PhysRevB.89.134302}. Soon after that, the VHE \cite{PhysRevLett.116.093901,PhysRevB.95.174106} and corresponding valley-protected edge states \cite{NatPhys.13.369,PhysRevB.97.155124,PhysRevApplied.9.034032,JAppPhys.123.091703,SciRep.7.10335,JApplPhys.123.091713} were theoretically predicted and experimentally observed in two-dimensional acoustic systems.

\begin{figure}[htbp]
\begin{center}
\includegraphics[width=0.8\columnwidth]{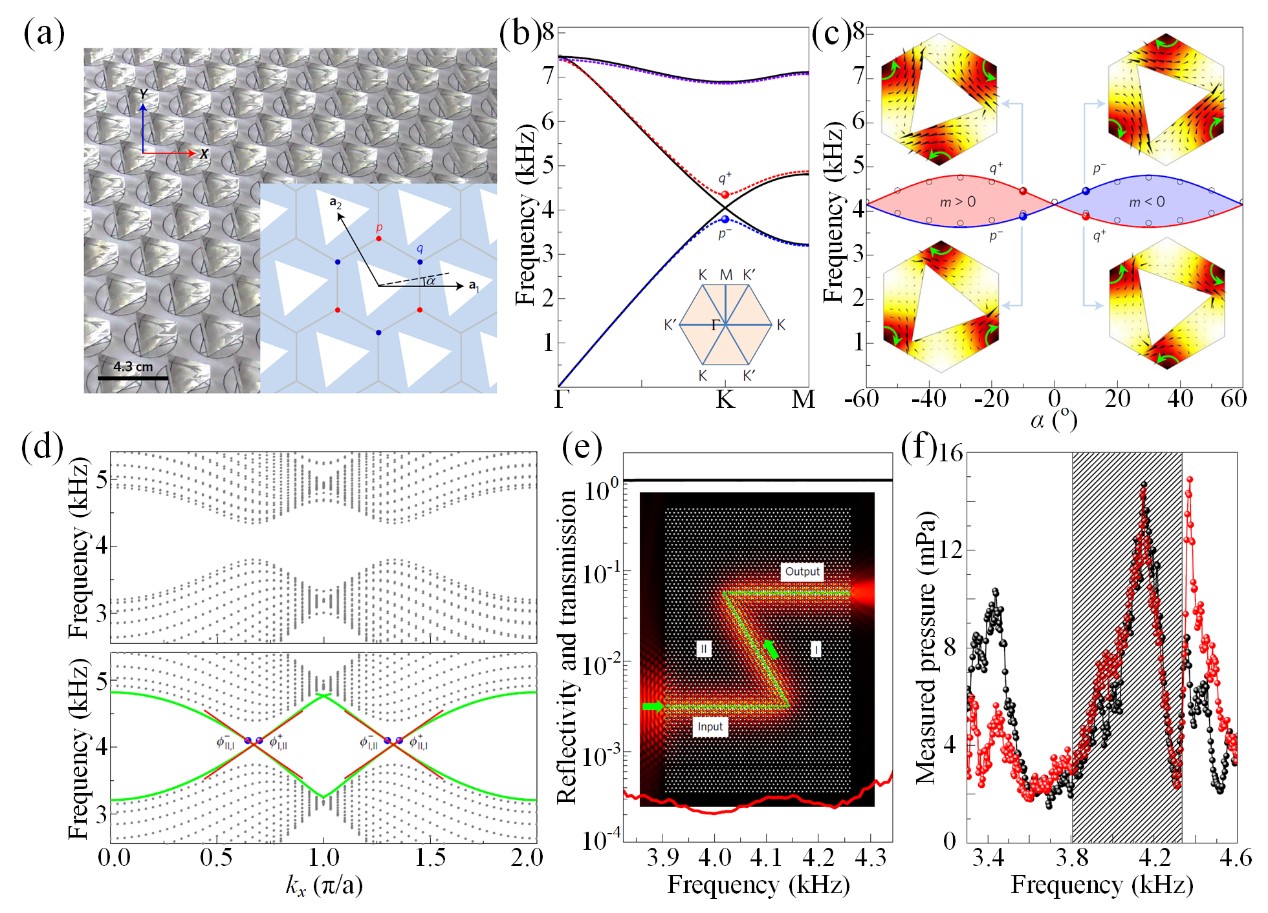}
\caption{Acoustic valley-Hall insulators. (a) The SC consisting of triangular polymethyl methacrylate rods positioned in a triangular lattice with lattice vectors $\mathbf{a_1}$ and $\mathbf{a_2}$. (b) Gapless and gapped bulk dispersion relations for the cases with $\alpha=0^\circ$ (black lines) and $-10^\circ$ (color lines). (c) Topological phase dependency with $\alpha$ (lines for simulations and circles for experiments), comprising specific vortex features (insets). (d) Comparative dispersions for the SC interfaces separating two topologically identical ($\alpha=10^\circ$ and $50^\circ$, upper panel) and separating two topologically distinct ($\alpha=-10^\circ$ and $10^\circ$, lower panel) AVH insulating phases, both simulated by a superlattice structure containing two different horizontal interfaces. (e) Power transmission (black line) and reflection (red line) spectra calculated for the two sharp turns in a zigzag path. Inset: field pattern. (f) Transmitted pressure measured for the zigzag path (red circles), compared with the one for a straight channel (black circles).}
\label{fig:AVHfig}
\end{center}
\end{figure}

Lu \emph{et al.} \cite{PhysRevLett.116.093901,NatPhys.13.369} firstly introduced the concept of valley states to SCs for acoustic waves. The hexagonal SC consists of triangular rods in a 2D waveguide, of which symmetries can be characterized by the rotation angle $\alpha$. It has been pointed out that the existence of a two-fold Dirac degeneracy at the corners of the 1st Brillouin zone (BZ) for any SC with $\alpha=n\pi/3$ is protected by the $C_{3v}$ symmetry, whereas the degeneracy would be lifted for any other rod orientation breaking the mirror symmetries \cite{PhysRevB.89.134302}. As shown in Fig. \ref{fig:AVHfig}(b), the dispersion relations for the SCs with $\alpha=0^\circ$ and $-10^\circ$ are illustrated. The vortex revolution at each valley (i.e., clockwise and anticlockwise) plays the role of the valley degree of freedom in a 2D acoustic system, as shown in insets of Fig. \ref{fig:AVHfig}(c). Figure \ref{fig:AVHfig}(c) shows the tuning of the AVH phase transition in a SC by variation of the rotating angle $\alpha$. When $\alpha<0^\circ$, the vortex chirality of the lower (upper) state is clockwise (anticlockwise), which appears exactly inverted when $\alpha>0^\circ$. The AVH phase transition that is accompanied by the crossing of these two pseudo spin states can be captured by an $\alpha$-dependent continuum Hamiltonian. Derived from the $\textbf{k}\cdot\textbf{p}$ theory, the unperturbed Hamiltonian ${\cal H}\left( {{k_ \bot }} \right) \equiv {{\cal H}_0}\left( {\delta k} \right)$ near the Dirac points can be written as \cite{PhysRevLett.95.226801,PseudoMagPNAS} ${{\cal H}_0}\left( {\delta k} \right) = {v_D}\left( {\delta {k_x}{\sigma _x} + \delta {k_y}{\sigma _y}} \right)$, where $v_{D}$ is the group velocity, $\delta k = \left( {\delta {k_x},\delta {k_y}} \right) \equiv {k_ \bot } - {k_D}$  is the distance from the Dirac points with  ${k_D} =  \pm \frac{{4\pi }}{{3{\alpha _0}}}{e_x}$ for the $K$ and ${K'}$ points, and ${\sigma _i}\left( {i = x,y} \right)$ are Pauli matrices of the vortex pseudo spins. The perturbation matrix is diagonalized: ${{\cal H}_P} = mv_D^2{\sigma_z}$. Consistent with the above band inversion picture (Fig. \ref{fig:AVHfig}(c)), the sign of the effective mass $m = {{\left( {{\omega _{{q^ + }}} - {\omega _{{q^ - }}}} \right)} \mathord{\left/
 {\vphantom {{\left( {{\omega _{{q^ + }}} - {\omega _{{q^ - }}}} \right)} {2v_D^2}}} \right.
 \kern-\nulldelimiterspace} {2v_D^2}}$ characterizes two different AVH insulators separated by the Dirac semi-metal phase with $m = 0$ in the phase diagram. The massive Dirac Hamiltonian $\delta H$ produces a nontrivial Berry curvature $\Omega \left( {\delta {\bf{k}}} \right) = \left( {{1 \mathord{\left/
 {\vphantom {1 2}} \right.
 \kern-\nulldelimiterspace} 2}} \right)m{v_D}{\left( {\delta {k^2} + {m^2}v_D^2} \right)^{ - {3 \mathord{\left/
 {\vphantom {3 2}} \right.
 \kern-\nulldelimiterspace} 2}}}$ in the first band. The integral of Berry curvature over the full BZ is zero with Chern number $C = 0$, which is required by time reversal symmetry. However, for small perturbation $m$, the Berry curvature is strongly peaked at the gap minima near $K$ and ${K'}$  points. The integral of the Berry curvature over an individual valley (one half of the BZ) is accurately defined and the non-vanishing valley-Chern indices can be determined by ${C_K} = {\mathop{\rm sgn}} {{\left( m \right)} \mathord{\left/
 {\vphantom {{\left( m \right)} 2}} \right.
 \kern-\nulldelimiterspace} 2}$ \cite{PNAS.110.10546,PhysRevB.88.161406}. As a result, the difference in the topological charge across the interface is quantized, which maintains a chiral edge mode according to the bulk-boundary correspondence \cite{PhysRevB.88.161406,PhysRevB.83.125109}. To verify this, two distinct SC interfaces are studied: one is constructed out of SCs with $\alpha  = {10^\circ }$ and $50^\circ$, of which the dispersion relation is illustrated in the top panel of Fig. \ref{fig:AVHfig}(d), and the other is constructed out of SCs with $\alpha  = {-10^\circ }$ and $10^\circ$, of which the dispersion relation is shown in the bottom panel. In the former case, the spectrum of relvance is completely gapped due to the presence of identical valley-Hall phases within the facing SCs. However, in the latter case, topological edge states fall within the bulk band gap as indicated by the green lines, which originate from AVH phase-inversion across the interface. Similar to the QSHE, edge states associated to the AVH effect appear robust against bends and crystal defects. Figure \ref{fig:AVHfig}(e) shows a negligible weak backscattering of the topological valley-projected edge mode propagating along an interface containg two sharp bends. The experimentally measured pressure amplitudes in the output channel of the interface with and without sharp bends are illustrated in Fig. \ref{fig:AVHfig}(f) demonstrating that the transmitted pressure of the edge states with bends almost entirely coincides with the one from a straight path within the topological band gap. Hence, although lattice-scale defects and obstacles may induce inter-valley scattering \cite{PhysRevApplied.9.034032,PseudoMagPNAS} with suppressed transmission, sharp bends as illustrated in Fig. \ref{fig:AVHfig}(e) indicate that backscattering-immune sound guiding in valley-Hall insulators is readily possible.

\begin{figure}[htbp]
\begin{center}
\includegraphics[width=0.8\columnwidth]{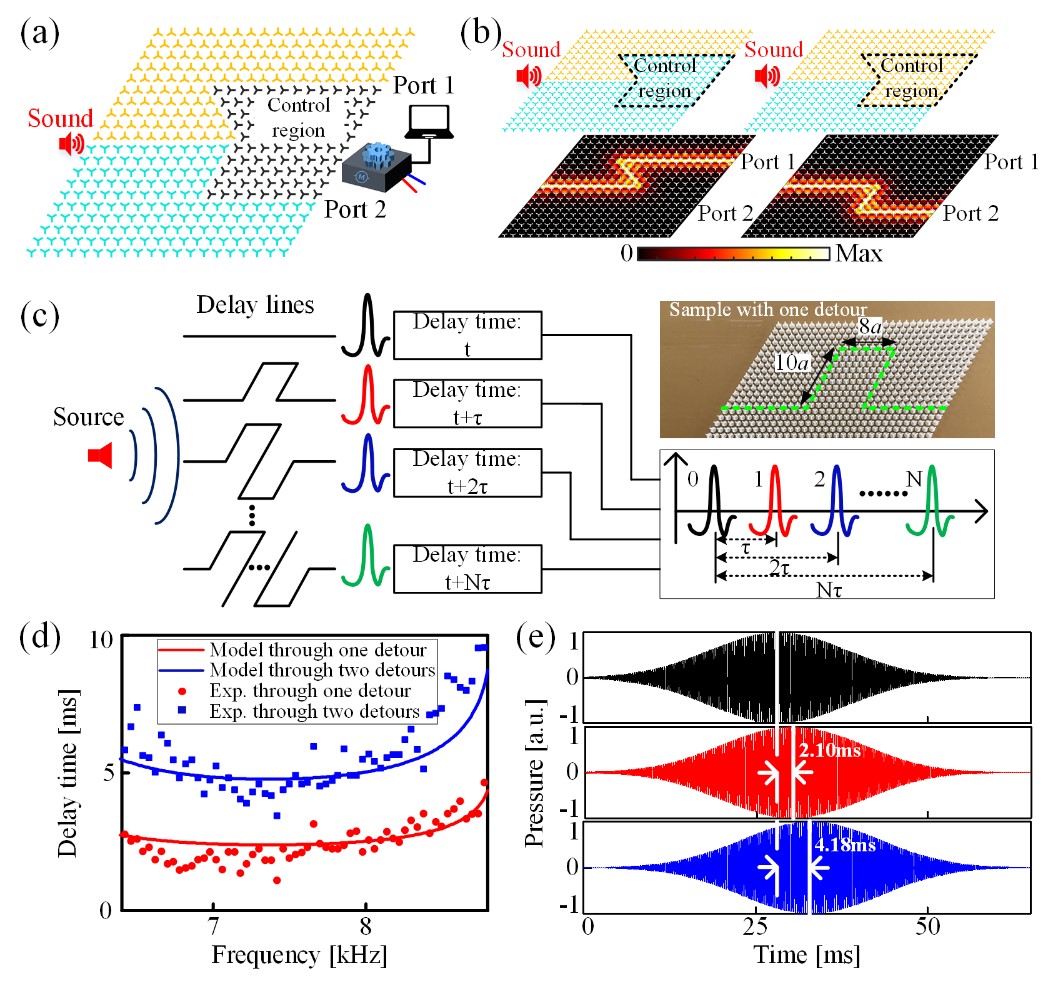}
\caption{Reconfigurable topological switch and broadband topological acoustic delay lines. (a) Setup of the reconfigurable topological switch. (b) The distributions of the pressure fields when switched between different topological states. (c) The schematic of time-delay lines based on the topologically protected waveguide. (d) The experimental time delays through one and two detours in the frequency range of the topological band gap, together with the model prediction. (e) Measured time delays through one detour (middle panel) and two detours (bottom panel), in which a Gaussian-modulated sinusoidal pulse is injected. The top panel shows the transmitted pressure through straight interface.}
\label{fig:TSandDLfig}
\end{center}
\end{figure}

As we mentioned earlier, the VHE enables exciting possibilities using the valley degree of freedom for valleytronics applications. Along the lines of technical implementation, a great challenge is posed by the lack of tunability and adaptation to functional needs concerning the AVH effect. Inspired by the proposal of delay lines in topological photonics \cite{hafezi2011robust,SciRep.6.28453}, Zhang \emph{et al.}\cite{PhysRevApplied.9.034032} experimentally realized topologically protected broadband delay lines based on engineered phase delay defects (PDDs) that constitute a new platform for acoustic devices. The structure consists of three-legged rods (TLRs) arranged into a triangular lattice providing an enlarged topological band gap through an optimized shape of these rods. The tunability of the unit cell is obtained through computer-controlled motors attached to the pedestals of the rods, which can provide the desired rotation. To further corroborate this approach and to demonstrate the actual proof-of-concept based on the reconfigurable topological edge states, the topologically switchable waveguides are designed as shown in Fig. \ref{fig:TSandDLfig}(a). Parts of the TLRs (highlighted by black color) are mounted onto computer-controlled motors. As a result, the topologically protected pathway can be easily controlled. Depending on the rotation angle selected, $-30^\circ$ or $30^\circ$, respectively, sound waves transmit either through Port $1$ or Port $2$ as illustrated in Fig. \ref{fig:TSandDLfig}(b). Vast possibilities of functional devices can be engineered by simply configuring the TLRs to any desired angle. Owing to the spectrally broad response of the TIs built using TLRs, a short pulses can be sent through them and their dynamic response can be readily engineered at will. The tremendous advantage of reflection-free acoustic signal transmission not yielding to sharp bends along the way, enables one to design acoustic delay lines by means of topologically protected transient edge states. Figure \ref{fig:TSandDLfig}(c) illustrates the PDDs in the form of square-shaped detours with four sharp bends along the interface that generates a time delay\cite{SciRep.6.28453}
\begin{equation}\label{eqn:timedelay}
{\tau _i}{\rm{ = }}\left( {\frac{{\partial {\varphi _i}}}{{\partial \omega }} - \frac{{\partial {\varphi _0}}}{{\partial \omega }}} \right){\rm{ = }}\frac{1}{{2\pi }}\left( {\frac{{\partial {\varphi _i}}}{{\partial f}} - \frac{{\partial {\varphi _0}}}{{\partial f}}} \right),\left( {i = 0,1,2 \ldots N} \right)
\end{equation}
where $\partial {\varphi _i}$  is the phase of sound wave along interface with $i = 0,1,2...N$ detours, and $\partial {\varphi _0}$ is the phase through the uninterrupted straight structure. The time delay can be increased very flexibly by successively stacking multiple square-shaped detours. According to the dispersion relation, the sound velocity can be obtained through  ${c_j} = 2\pi  \cdot {\left( {{{df} \mathord{\left/
 {\vphantom {{df} {dk}}} \right.
 \kern-\nulldelimiterspace} {dk}}} \right)_j}$ with $j = 1, 2$ representing different kind of interfaces. The delay time $\tau {\rm{ = }}\sum\limits_{j = 1}^2 {\frac{{{s_j}}}{{{c_j}}}}$ can be derived from Eq.(\ref{eqn:timedelay}), where ${s_j}$ is the length of the transmitted line. The time delays of sound waves through a single and a double PDD are experimentally measured as shown in Fig. \ref{fig:TSandDLfig}(d) together with the model prediction. Figure \ref{fig:TSandDLfig}(e) depicts the experimentally measured time delays through interfaces containing the aforementioned PDDs. These experimental results constitute the first step in building a multi-stage broadband topologically protected delay line capable of buffering multiple acoustic pulses.

\begin{figure}[htbp]
\begin{center}
\includegraphics[width=0.8\columnwidth]{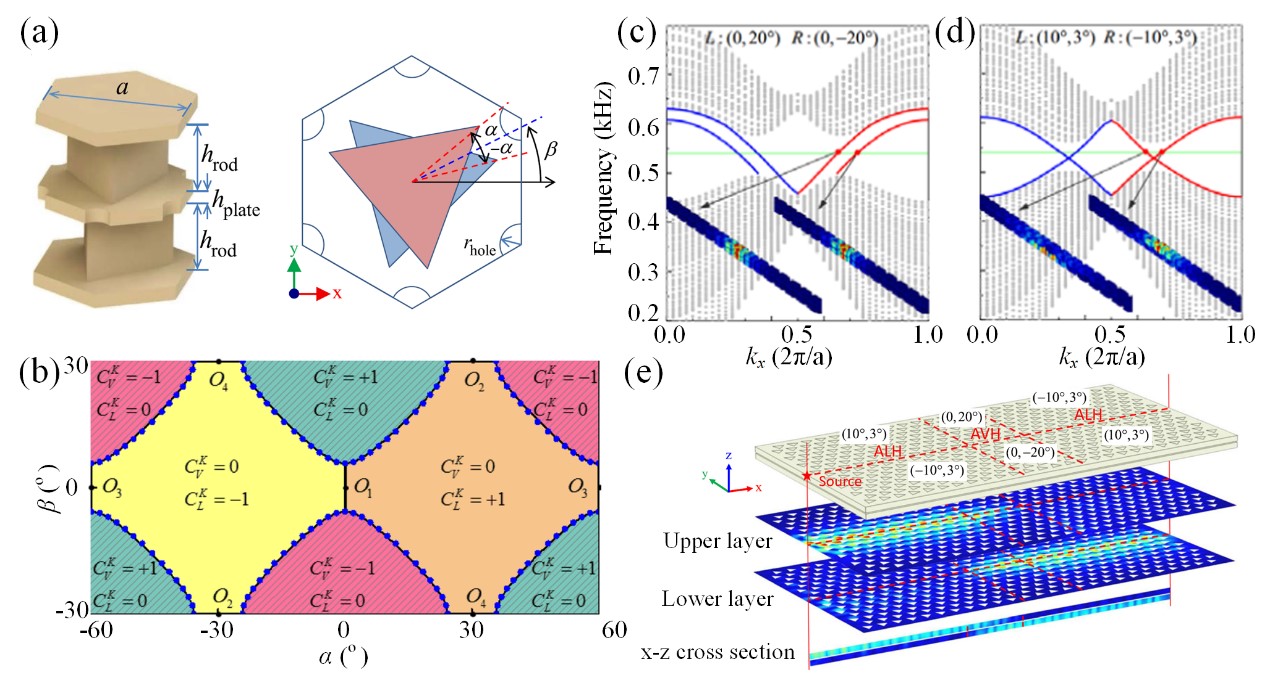}
\caption{Valley topological phase in a BSC. (a) Side (left) and top (right) views of the unit cell. (b) Reduced phase diagram parametrized by the angles ($\alpha$, $\beta$). The numerical phase boundaries (solid lines) correspond to the closure of the omnidirectional band gap, consistent with the model predictions (dots). Projected dispersion relation along an interface separating two topologically distinct AVH BSCs (c), and two topologically
distinct ALH BSCs (d). (e) Full-wave simulation at 15.3 kHz, which shows sound concentration varying from the upper layer to the lower layer.}
\label{fig:bilayerfig}
\end{center}
\end{figure}

Beyond valley-Hall phases within a single SC, layer-mixed and layer-polarized topological valley Hall phases were recently proposed by Lu \emph{et al} \cite{PhysRevLett.120.116802}, using a unique design of bilayer sonic crystals (BSCs). Figure \ref{fig:bilayerfig}(a) illustrates the schematic of the unit cell of the BSC, which is composed of two layers of finite SCs sandwiched between a pair of rigid plates and separated by a plate that is pierced with a honeycomb array of circular holes. Each layer consists of a hexagonal array of regular triangular rods. As a result, the degree of freedoms of orientations is broad with respect to the relative angle $\alpha$ and the common angle $\beta$. Figure \ref{fig:bilayerfig}(b) depicts a phase diagram in the entire angular domain, illustrating straight and curved lines, which are associated with ring and point degeneracies, respectively. The shaded regions represent the nontrivial AVH phases that are characterized by the quantized topological invariant $C_V^K$  associated to the single-layer system, whereas the remaining regions of the phase diagram represent nontrivial acoustic layer-valley Hall (ALH) phases, characterized by $C_L^K$. To observe the differences between those phase, dispersion relations of ribbon-shaped BSCs are studied: first, a BSCs is composed with rod orientations $\left( {0, \pm 20^\circ } \right)$, which belong to different AVH phases ($\Delta C_V^K = 2$). As shown in Fig. \ref{fig:bilayerfig}(c), two edge modes with positive group velocities appear at the $K$ valley (red curves) and two other edge modes with opposite group velocities (blue curves) appear at the ${K'}$ valley due to the time-reversal symmetry. The eigenfields are located in both layers (see insets). Second, a BSCs is composed with rod orientations $\left( { \pm 10^\circ ,3^\circ } \right)$, which belong to different ALH phases ($[\Delta C_L^K = 2$). Compared to the AVH scenario, the group velocities of the two edge modes at the K valley are opposite. The eigenfield concentrates predominantly in either the upper layer or the lower one. That is to say, for the edge state projected by the same valley, sound waves propagate towards one side in the upper layer and towards another side in the lower layer without any interference with each other, which is equivalent to spin-orbital couplings in electronic systems. As shown in Fig. \ref{fig:bilayerfig}(e), an efficient inter-layer converter can be constructed by four distinct BSC phases that support ALH (bilateral) and AVH (middle) edge modes along the $x$ direction interfaces. When a point source is placed at the left side of the upper layer, most of the sound energy is switched to the lower layer as the wave reaches another ALH interface, assisted by the layer-mixed AVH interface with specific length.

\section{Three dimensional topological acoustics}

Parallel to developments of TIs, topological semimetals have emerged as a new frontier in the quest of new topological phases in the past few years. \cite{Burkov1,chenfang2}  Topological semimetals are identified as topological materials in the sense that the gapless band structures are topologically protected and are accompanied by robust gapless surface states. \cite{Burkov2} Weyl semimetal, as an important member of the topological semimetals, has received quite a bit of attention recently with the theoretical discovery and experimental realizations in electronic and photonic systems. \cite{Fang92,Burkov3,xiangangwan1,Suyangxu1,BQLv1,Luling1,Luling2} Soon after that, acoustic Weyl metacrystals were discovered in coupled resonators and waveguides \cite{MX1,ZhaojuYang1} and experimentally demonstrated very recently \cite{FengLi1,Minghuilu2018,BaizhanXia2018}. In addition to airborne sound waves\cite{MX1,ZhaojuYang1,FengLi1,Minghuilu2018,BaizhanXia2018}, Weyl points have also been investigated in elastic waves \cite{YaotingWang2018,Fruchart201720828}. To illustrate the physics, we focus on airborne sound waves.

Weyl semimetals are periodic systems that possess Weyl points \cite{Weyl1929}, which are topological robust band degeneracy points. The  Hamiltonian which describes a Weyl point with a topological charge of +1 or $-1$ is given by
\begin{equation}
\hat{H}(\boldsymbol{q})= f(\boldsymbol{q})\sigma_0+\sum_{i,j=x,y,z} q_i v_{i,j} \sigma_j,
\label{eqn:WeylHmt}
\end{equation}
where $\boldsymbol{q}=(q_x,q_y,q_z)$ and $q_i$ is the wave vector originating from the Weyl point, $f(\boldsymbol{q})$ is an arbitrary real function of $\boldsymbol{q}$, $ \boldsymbol{v} $ is a $3 \times 3$ constant matrix, $\sigma_0$ is the $2\times 2$ unit matrix, and $\sigma_x, \sigma_y, \sigma_z $ are the Pauli matrices. The charge of the Weyl point in Eq.( \ref{eqn:WeylHmt}) is given by $C = \textnormal{sgn} (\det \boldsymbol{v}) $.  Weyl points can also exhibit higher topological charges. \cite{chenfang1} For simplicity sake, we restrict our discussion here  to Weyl points of charge +1 and $-1$. Weyl points are robust against any perturbations which keep the wave vectors as good quantum numbers. This can also been seen from the fact that the Weyl Hamiltonian contains all the Pauli matrices and hence local perturbation can only serve to shift the Weyl point in momentum space but cannot open a band gap. A Weyl point can only be "annihilated" by another Weyl point which carries the opposite topological charge and a band gap can be opened if Weyl points carrying opposite charges collide in momentum space.

\begin{figure}[htbp]
\begin{center}
\includegraphics[width=1.0\columnwidth]{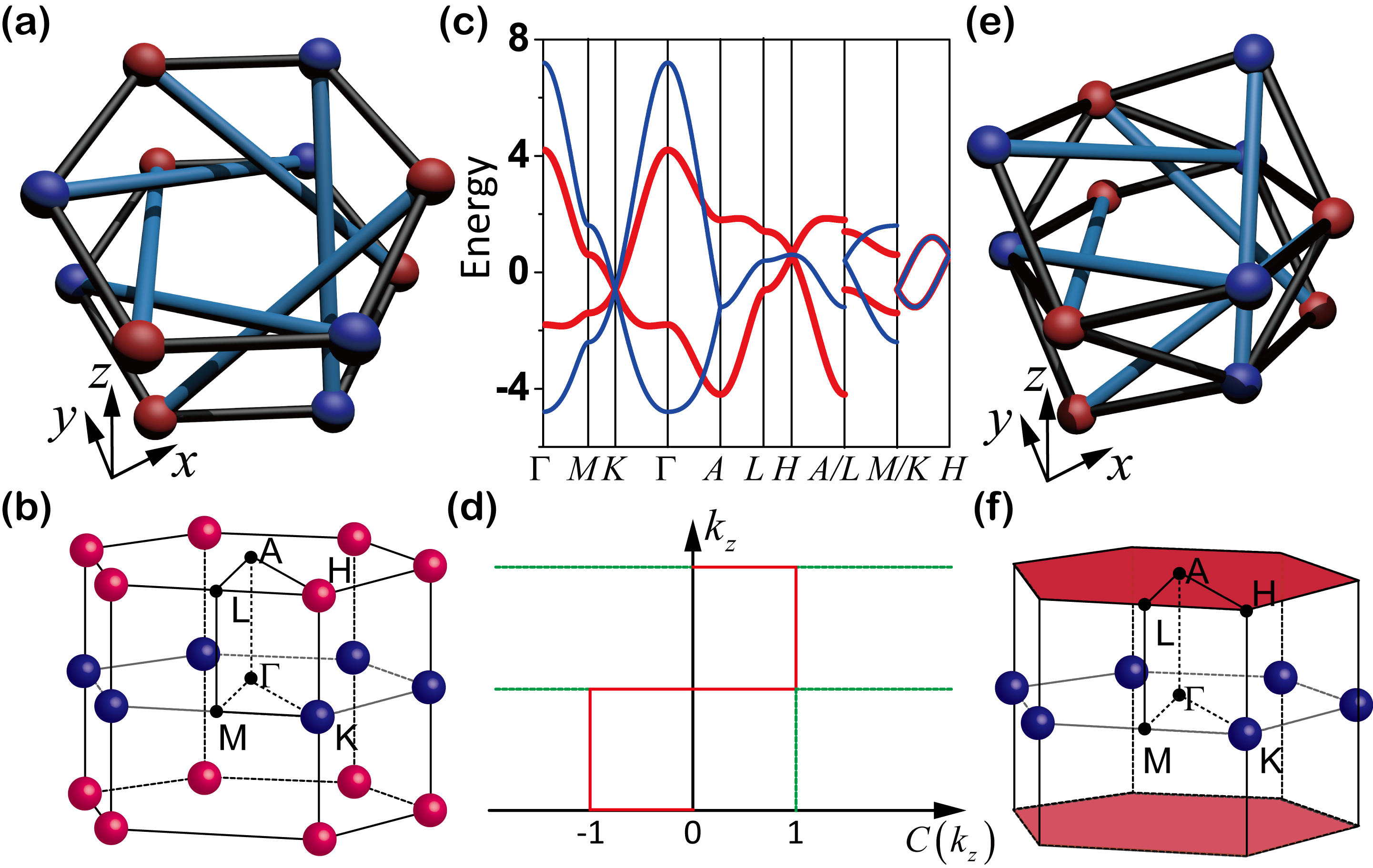}
\caption{Weyl semimetal and topological charged nodal surface semimetal. (a) and (e) show the tight binding model of a system possessing Weyl points and a charged nodal surface, respectively. (b) and (f) show the topological charge distribution in the reciprocal space for the tight binding model in (a) and (e), respectively, where the red and blue spheres represent Weyl points with charge +1 and $-1$, respectively, and the red surface represents a nodal surface with topological charge +2. Some high symmetry points in the first Brillouin zone are also labeled. (c) shows the band structure, where $t_0=1$ and $t_c=0.2$ are used. Here red and blue correspond to the tight binding model in (a) and (e), respectively. (d) shows the Chern number as a function of $k_z$, which changes at $k_z=0$ and $k_z=\pi$. }
\label{fig:Weyl1}
\end{center}
\end{figure}

Acoustic Weyl metacrystals were first discussed in 2015 in a system composing of coupled resonators and coupled waveguides. \cite{MX1}  Figure \ref{fig:Weyl1}(a) shows the schematic picture of a tight binding model that explains the formation Weyl points in the reciprocal space for that particular system. This tight binding model can be regarded as an AA stacking of graphene lattice along the $z$ direction, where the red and blue spheres represent different sublattices and the black bonds represent intralayer hopping with hopping strength $t_0$. Interlayer hopping is represented by the cyan bonds and with hopping strength $t_c$. In this tight binding model, time reversal symmetry is preserved and hence both $t_0$ and $t_c$ are real constants. The Hamiltonian $\hat{H} $ of this tight binding model is given by:
\begin{equation}
\hat{H}= \sum_{\langle i,j \rangle} t_0 b_{i,k}^{\dagger}a_{j,k}+\sum_{\langle\langle i,j\rangle\rangle}t_c (a_{i,k+1}^{\dagger} a_{j,k}+b_{i,k+1}^{\dagger} b_{j,k})+H.c.,
\label{eqn:Weyltb}
\end{equation}
where $a (b) $ and $a^\dagger (b^\dagger) $ are the annihilation and creation operators on the sublattice cites. Each lattice site is labeled by $i,k$, wherein the first denotes the coordinate inside each layer and the second denotes the layer number. $\langle i,j \rangle $ in the first summation represents intralayer nearest neighbor and $\langle\langle i,j \rangle\rangle $ in the second summation represents interlayer next nearest neighbor. The corresponding Bloch Hamiltonian $H(\boldsymbol{k}) $ is given by
\begin{equation}
\hat{H}(\boldsymbol{k})=
\begin{pmatrix}
t_c g(k_z) & t_0 h(k_x,k_y) \\
t_0 [h(k_x,k_y)]^* & t_c g(-k_z)
\end{pmatrix}
\quad
,
\label{eqn:Weyltbb}
\end{equation}
where $ g(\boldsymbol{k})=2\cos(k_xa+k_zd_h)+4\cos(\sqrt{3} k_ya/2)\cos(k_xa/2-k_zd_h) $ and $h(k_x,k_y)=\exp(-i\sqrt{3}k_ya/3)+2\cos(k_xa/2)\exp(i\sqrt{3}k_ya/6)$   with $a$ and $d_h$ being the lattice constants of the graphene lattice and along the $z$ direction, respectively.

The first Brillouin zone of this tight binding model is shown in Fig. \ref{fig:Weyl1}(b) with some high symmetry points also being labeled. The band structures along some high symmetric directions are shown in Fig. \ref{fig:Weyl1}(c) with the red curve, where the hopping parameters are taken to be $t_0=1$ and $t_c=0.2$. The band dispersions are linear near the points $K$ and $H$ along all the high symmetric directions, which indicates that this tight binding model possesses Weyl points at $K$ and $H$ as well as $K'$ and $H'$. To see this point, we expand the Hamiltonian in Eq. (\ref{eqn:Weyltb}) around $K=(0, 4\pi a/3, 0)$ . Keeping only to the lowest order, we get
\begin{equation}
\hat{H}(\boldsymbol{q})= -3t_c\sigma_0+3\sqrt{3}t_cq_z\sigma_z-\dfrac{\sqrt{3}}{2}t_0(q_y\sigma_y-q_x\sigma_x).
\label{eqn:Weylexp}
\end{equation}
The first term in Eq. (\ref{eqn:Weylexp}) represents an energy shift which does not change the topological charge of the Weyl point. And hence the charge of the Weyl point at $K$ is $-1$. The system exhibits $C_6$ rotational symmetry along the $z$ direction, and time-reversal symmetry implies that the system possesses another Weyl point at the $K'$ point and with the same charge. One can also follow the same procedure and concludes that the charge of the Weyl point at $H$ is +1. This fact also illustrates an important property of Weyl crystals: the total charge inside the first Brillouin zone of a periodic system should vanish. \cite{NIELSEN1983389} This distribution of the Weyl points and their associated topological charges are shown in Fig. \ref{fig:Weyl1}(b).

In this system, $k_z $ is a good quantum number, and hence one can define the Chern number of any two-dimensional subsystems with a fixed $k_z$. In Fig. \ref{fig:Weyl1}(d), we show the Chern number as a function of $k_z$ with the red curve. The Chern number only changes when it comes across a $k_z$ plane with nonzero topological charge. In this system, the Chern number is +1 for $k_z>0$ and $-1$ for $k_z<0$. According to the bulk-edge correspondence \cite{Hatsugai1}, there should be one-way edge states provided that $k_z$ is preserved. Such one-way edge states should be present for any $k_z\neq 0 $ or $ \pi$ which then forms a  Riemann surface like structure and the iso-energy contour of it is known as the Fermi arc.\cite{chenfang2}

\begin{figure}[htbp]
\begin{center}
\includegraphics[width=1.0\columnwidth]{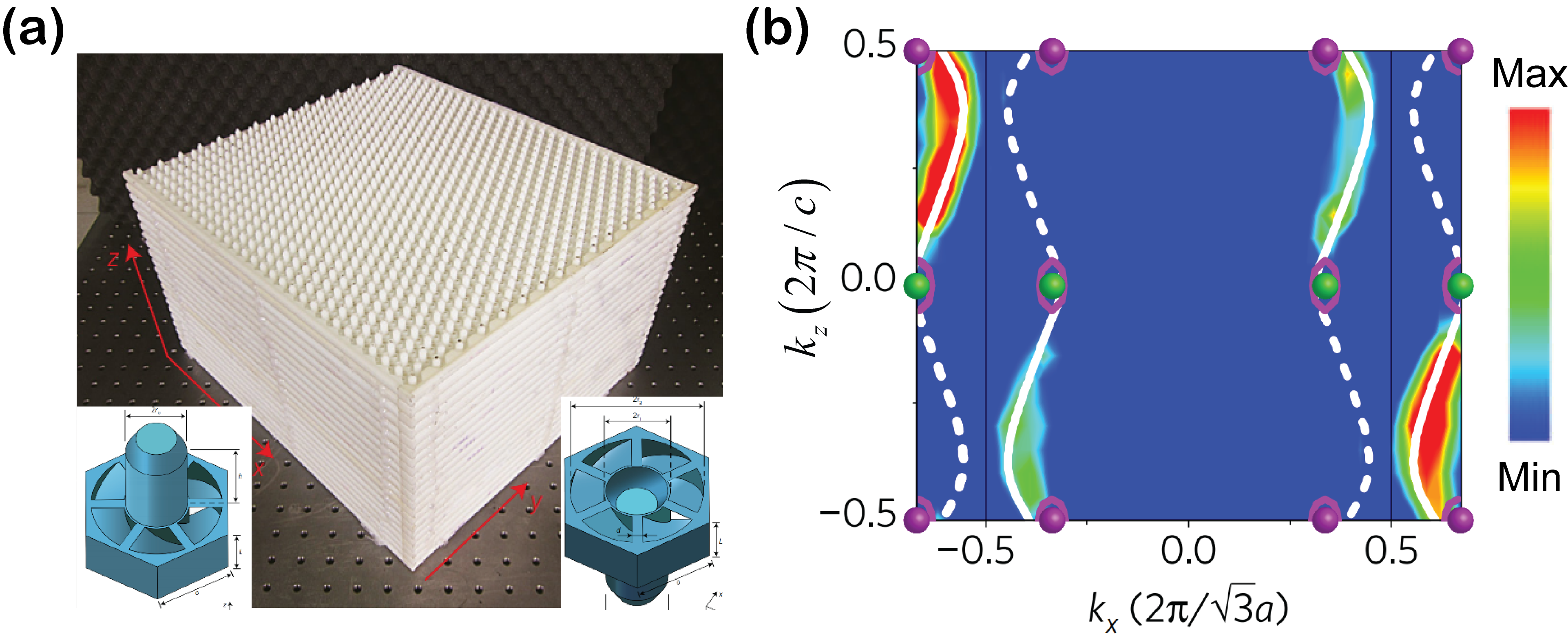}
\caption{Experimental observation of acoustic Fermi-arcs in Ref.\cite{FengLi1} (a) A photo of the sample, where the insets show the front (lower left) and back (lower right) views of the unit cell. (b) Fermi-arcs measured using Fourier transforms of surface wave fields. }
\label{fig:Weyl2}
\end{center}
\end{figure}

One advantage of using acoustic metamaterial as a platform to explore the physics of topological material is that real samples can be made more straightforwardly than quantum materials. Indeed, experimental realizations of acoustic Weyl metacrystals were performed soon after the theoretical proposal. \cite{FengLi1,Minghuilu2018,BaizhanXia2018} We focus on the experiments in Ref. \cite{FengLi1}. In this work, the authors realized an acoustic Weyl metacrystal, measured the Fermi arc and demonstrated the robustness of the surface states. Figure \ref{fig:Weyl2}(a) shows the experimental sample and the insets show the front (lower left) and back (lower right ) views of the unit cell. The sample was fabricated by 3D printing technology, and the printed structure can be treated as sound hard where the sound waves cannot penetrate. The detail geometric parameters can be found in Ref. \cite{FengLi1}. In this system, the Weyl points with opposite charges are not related by any symmetry and hence their frequencies can be different. The frequencies of the Weyl points under consideration at $K$ and $H$ are 15kHz and 16kHz, respectively. Figure \ref{fig:Weyl2}(b) shows the surface Brillouin zone on the $k_x-k_z$ plane and projection of Weyl points as denoted by the green and purple spheres. The solid and dashed curves represent the "Fermi arcs" on the positive and negative $x-z$ surfaces, respectively. The working frequency is chosen to be 15.4kHz which is between the frequencies of these two Weyl points. At this frequency, the equi-frequency contours of the bulk bands around the Weyl points project to elliptical disks as outlined by the purple curves. The Fermi arcs connect these elliptical disks. Experimentally, one can measure the surface wave field distributions which can be Fourier transformed to obtain the Fermi arc. The experimental results are shown as color code in Fig. \ref{fig:Weyl2}(b), where red represents maximal value and blue represents minimal value. The experimental measured results agree quite well with the numerical simulations. As discussed before, such systems support one-way surface states against $k_z$ preserved scatterings, which are also experimentally demonstrated. \cite{FengLi1,Minghuilu2018,BaizhanXia2018}

Weyl points are not the only object in the reciprocal space that possesses a topological charges \cite{MX2}. A nodal surface, which is a surface degeneracy between two bands, can also possess a nonzero topological charge. The nodal surface is protected by non-symmorphic symmetry $G_{2z}\equiv \mathcal{T} \tilde{C}_{2z} $ and is located at $k_z = \pi/d_h $ with arbitrary $k_x $ and $k_y$, where $\mathcal{T}$ denotes the time reversal operation and $\tilde{C}_{2z}$ represents the two-fold screw rotational symmetry along the $z$ direction. The band dispersion is linear in the vicinity of the nodal surface. Note here that $G_{2z}$ symmetry only protects the presence of the nodal surface, whether it is topologically charged or not depends on the system parameters. \cite{MX2} A tight-binding model that exhibits this symmetry and also possesses a charged nodal surface can be obtained by simply shifting one of the sublattice in Fig. \ref{fig:Weyl1}(a) along the $z$ direction by $d_h/2$, and the resulting tight binding model is shown in Fig. \ref{fig:Weyl1}(e). Here, the intralayer hopping is also slightly modified to preserve the $G_{2z}$ symmetry. The band structure of this tight binding model is shown in Fig. \ref{fig:Weyl1}(c) with the blue curve. Here the intralayer hopping (denoted by the black bonds) and interlayer hopping (denoted by the cyan bonds) are set as $t_0=1$ and $t_c=0.2$, respectively. We see that the two bands become degenerate on the $k_z=\pi$ plane and the dispersion is linear away from this nodal surface. The Weyl points at $K$ and $K'$ still are preserved while the Weyl points at $H$ and $H'$ are merged into the nodal surface and hence the nodal surface should possess topological charge +2. The charge distribution of the tight binding model in Fig. \ref{fig:Weyl1}(e) is shown in Fig. \ref{fig:Weyl1}(f), where the red plane represents the nodal surface with topological charge +2. The Chern numbers as a function of $k_z$ remain the same as that in Fig. \ref{fig:Weyl1}(d).

\section{Topological mechanical waves}

In the preceding sections we reviewed the entire landscape of topological states in both time reversal symmetric and asymmetric acoustic structures. It is thus evident that these quantum topological phenomena, similar to photonic systems, find their counterparts in engineered sonic structures and lattices. But does this apply to mechanical waves as well? Throughout this review, we have hinted towards it, first of all however, we need to distinguish between zero frequency nontrivial topological modes that are insensitive to smooth deformations and actual mechanical vibrations. In this section we embark to discuss on the latter whereas topological zero modes and states of self-stress have been extensively discussed elsewhere \citep{MechFlex}.The study into topologically protected mechanical waves at finite frequencies has been largely triggered by the search of related phenomena for other areas of classical physics \cite{PhysRevX.5.031011,PhysRevB.96.064106}. Hence, for elastic vibrations and mechanical waves, equivalent to the preceding sections we discriminate between systems of intact and broken time-reversal symmetry, which will ultimately give rise interface-supporting modes in the form of helical spin-polarized and chiral one-way edge states, respectively. Gyroscopic lattices are prominent candidates capable to form forbidden regions, band gaps, in which bulk vibrations are inhibited. The opening of such gap is induced by breaking the $\CMcal{T}$ symmetry through rotating gyroscopes permitting waves to sustain only in the form of unidirectional and topological robust chiral edge states along finite sample interfaces \cite{BertoldiGyro,VitelliGyro}.

In order to mimic with mechanical waves the QSHE and launching helical edge vibrations at interfaces one can revert to passive systems, i.e. $\CMcal{T}$ symmetric structures in no need for an external bias to violate reciprocity. In order to keep the $\CMcal{T}$ symmetry intact, two opposite counter-propagating spins have to coexist at the same frequency according to Kramers degeneracy theorem.  Hence, helical edge states have been detected in pendula lattices \cite{SusstrunkSCIENCE,Hexaspring} and predicted for Lamb and flexural waves in structured plates \cite{NatComm2015,torrentLike} and granular crystals\cite{Tournat}.

As detailed above, in addition to the pseudo spin, sound waves are capable of emulating a valley degree of freedom. Rather than pursuing novel carriers of information and energy similar to their electronic counterpart, recent efforts in the literature have demonstrated that valley Hall polarized mechanical states can give rise to topological protected vibrations. In order to gap the Dirac cones of phononic lattices, the inversion symmetries were broken by introducing a hexagonal boron nitride-like geometry of unequal masses within the unit cell. It was shown that these elastic systems contain topologically nontrivial bandgaps hosting backscattering suppressed edge states\cite{RuzzeneNJPhBN,RuzzeneHexagonalPRB}. Similar efforts have been devoted to observe topologically valley-polarized states in bilayers, slender veins connected rod-crystals, thin plates, and diatomic waveguides\cite{RuzzeneBilayer,SciRep.7.10335,ValleyLamb,ValleyexpPRB}.

Recently, on the nanoscale, chiral phonons were observed in monolayer tungsten diselenide whose broken $\CMcal{P}$ symmetry splits clockwise and counterclockwise motions into nondegenerate states, which has great potential for the realization of phonon-driven topological states and controlled intervalley scattering \cite{ChiralPhononsZhang}. These topological states are based on atomic-scale lattice vibrations that differ substantially from the previously mentioned states in macroscopical artificial crystals with lattice constants in the \textit{mm} to \textit{cm} range. Along the same frontier, Weyl phonons have been predicted in magnetic- and transition metals. In the latter case, double Weyl points are hosted throughout the phonon spectrum thanks to the noncentrosymmetric but $\CMcal{T}$ symmetric crystalline structure \cite{DoublePhononWeyl}. In contrast, in the former case, it was shown that strong circular phonon dichroism can be induced in Weyl semimetals with both of broken $\CMcal{P}$ and $\CMcal{T}$ symmetry \cite{PhononWeyl}. These findings have the possibility to open doors for topological phononics at the atomic scale comprising engineering surface phonons and heat transport related to the phonon Hall effect \citep{PhysRevLett.105.225901}.

Other exiting explorations along the mechanical research frontier of topological insulators show how smooth deformations can provide topological phase transitions, pseudomagnetism for sound is created, and how mechanical quadrupole topological insulators sustain corner states \cite{SoftNatComm,PseudoMagPNAS,Quatru}. Conclusively, these tremendous efforts demonstrate that man-made artificial crystals serve as a fruitful playground to test with mechanical vibrations topologically protected wave propagation.

\section{Future directions}
We have reviewed recent efforts in putting topological acoustics on the map of an overall attempt in building the bridge between quantum physics and topological insulators for classical waves. Artificial macroscopic lattices in the form of sonic and phononic crystals pose only little fabrication challenges as compared to photonic systems that are built on much smaller scales at which imperfections matter. Concerning this, it did not take much time for the frontier of topological acoustics to flourish from 2D to 3D where theoretical predictions quickly turned into experimental proofs of concepts.

The strategy to move forward could potentially rest on looking into the current state-of-the-art comprising electronic and photonic topological systems. Doing this will certainly provide a platform for research into exotic sound propagation and mechanical vibrations to spearhead novel basic wave physics. Topics such as higher-order topological insulators (HOTI)\cite{Quatru} sustaining corner and hinge states, experimental progress in non-Hermitian and $\CMcal{PT}$ symmetry topological insulators \cite{PhysRevLett.120.246601}, and analogies of Majorana-like bound states at finite frequencies will fall into this category making topological acoustics relevant among condensed-matter physicists. However, it is of substantial significance to actually look at the specific targets of research in acoustics and elasticity together with their more technological oriented challenges. Realizing topological robust and defect insensitive wave guiding, signal buffering and splitting could provide new avenues for improved surface acoustic waves sensors, on-chip filters in mobile phones, enhanced coupling efficiency in touchscreens, and potentially improve the robustness of bio-chemical sensing. Along this line, efforts in the near future must therefore revolve around shrinking topological acoustical but also mechanical properties into the micro scale in order to be of relevance to the aforementioned technologies.

Conclusively, while many of the reviewed results and breakthrough findings for the most part can be categorized in terms of academic research, based on the unconventional way sound and vibrations are tailored, one can only look forward to unprecedented routes and possibilities for phononic technologies that fully take advantage of topologically robust wave control.

\newpage
%\bibliographystyle{naturemag}
%\bibliography{Bibliography}

\end{document}